# RADAR: Recall Augmentation through Deferred Asynchronous Retrieval

Amit Jaspal
Meta Platforms, Inc.
Menlo Park, CA, USA
ajaspal@meta.com

Qian Dang
Meta Platforms, Inc.
Menlo Park, CA, USA
qdang@meta.com

Ajantha Ramineni
Meta Platforms, Inc.
Menlo Park, CA, USA
aramineni@meta.com

## ABSTRACT

Modern large-scale recommender systems employ multi-stage ranking funnel (Retrieval, Pre-ranking, Ranking) to balance engagement and computational constraints (latency, CPU). However, the initial retrieval stage, often relying on efficient but less precise methods like K-Nearest Neighbors (KNN), struggles to effectively surface the most engaging items from billion-scale catalogs, particularly distinguishing highly relevant and engaging candidates from merely relevant ones. We introduce Recall Augmentation through Deferred Asynchronous Retrieval (RADAR), a novel framework that leverages asynchronous, offline computation to pre-rank a significantly larger candidate set for users using the full complexity ranking model. These top-ranked items are stored and utilized as a high-quality retrieval source during online inference, bypassing online retrieval and pre-ranking stages for these candidates. We demonstrate through offline experiments that RADAR significantly boosts recall (2X Recall@200 vs DNN retrieval baseline) by effectively combining a larger retrieved candidate set with a more powerful ranking model. Online A/B tests confirm a +0.8% lift in topline engagement metrics, validating RADAR as a practical and effective method to improve recommendation quality under strict online serving constraints.

## 1  Introduction

Modern video-sharing platforms confront an extreme retrieval challenge: every user session must search through billions of candidate videos to prepare a personalized list within a few milliseconds. Production systems therefore follow a three-stage funnel — retrieval → pre-ranking → ranking — where a very lightweight retriever supplies roughly $O(10^3)$ items, a moderate pre-ranker trims this set to the low hundreds, and an expressive ranker finally orders the shortlist for display. However, this cascaded design imposes hard ceiling on recall because the retriever must satisfy the tightest latency budget, it relies on inexpensive signals (e.g. dot-product two-tower models [1, 2] or K-nearest-neighbour CF indices) and consequently fails to surface many highly engaging items. In fact, in offline simulation studies, we observe single-digit Recall@200 from standard user-to-item and item-to-item retrieval methods (Fig. 1). This results in a retrieval bottleneck that throttles downstream ranking quality.

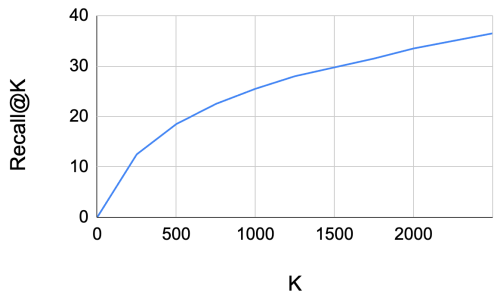

Figure 1: Recall@K as a Function of Candidate Retrieved Size (K) from Two Tower based DNN model

Research has therefore explored hybrid offline–online architectures that pre-compute richer candidates when latency is less constrained. These often focus on pre-computing sophisticated retrieval candidates using complex offline models, subsequently relying on lightweight online models for serving [5, 6]. For instance, TwERC [6] augments a real-time lightweight ranker with complementary sources like graph-based neighbors and cached ranker scores, significantly improving coverage. Similarly, other hybrid architectures combine batch-trained models with real-time bandit layers [5] to balance exploration and exploitation. Recent papers also revisit retrieval using accelerator-backed deep matching models to close the expressiveness gap between retrieval and ranking [7].

While these approaches offer valuable improvements, they still restrict the retrieved candidates to what a server-side model can compute in real time. We observe that significant computational resources are frequently available during off-peak hours. This observation motivates our proposed Recall Augmentation through Deferred Asynchronous Retrieval (RADAR), a novel framework that leverages these off-peak resources **to** perform computationally expensive final-stage ranking step on a much larger set of candidates (50X) asynchronously, before the user session begins.

To the best of our knowledge, no prior hybrid retrieval system—including TwERC [6]—fully decouples candidate generation from online serving **constraints** by (i) running the production-grade ranking model offline on a 50X larger set of retrieved candidates and (ii) refreshing each user's pre-ranked list on a usage-adaptive cadence

## 2  Proposed Approach - RADAR

## 2.1 System Overview

Figure 2 illustrates the RADAR architecture. The online path handles real-time user requests, fetching candidates from standard retrieval sources and the newly introduced RADAR key-value store. Candidates from standard sources pass through pre-ranking, while RADAR candidates bypass it. All candidates are then merged and ranked by the final online ranking model. The offline path runs asynchronously—scheduling, generating and refreshing the RADAR key-value store.

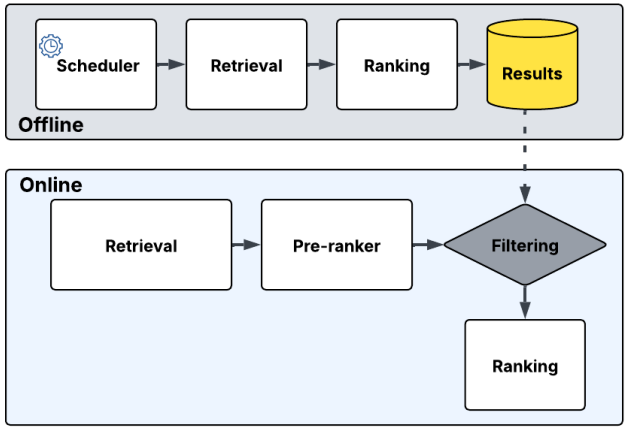

Figure 2: RADAR End-to-End System Architecture

## 2.2 Offline RADAR Pipeline

This pipeline runs periodically to generate high-quality candidates for users

1. **Triggering & Scheduling**: The pipeline is triggered based on user activity. For instance, active users might have their lists refreshed daily, while less active users might be refreshed weekly or bi-weekly. This usage-based scheduling ensures top K results are updated consistently while managing computational costs. The jobs run on elastic, preemptible compute resources during off-peak hours to minimize operational cost. Note that even with usage-based trigger, scheduling will happen only during off-peak window, hence users key-value store results can be stale for few hours in worst case.
2. **Large-Scale Candidate Generation**: For each targeted user, we perform retrieval using existing, efficient methods (like Two Tower DNN, Content KNN, ItemCF and other rule-based sources) but configure them to retrieve a significantly larger number of candidates (~ 50X the online output size). This broad set aims to capture a diverse range of potentially relevant items that might be missed by the constrained online retrieval.
3. **Offline Ranking**: This is the core of RADAR, the retrieved 50X candidates for each user are scored using the exact same complex, feature-rich ranking model used in the final stage of the online funnel. Critically, this allows us to apply our best predictive model to a vastly larger set than feasible online.
4. **Storage**: The top 200 highest-scoring items, along with their ranking scores, are stored per user in a low-latency key-value store.

Note that the offline RADAR pipeline operates on an opportunistic compute tier during off-peak periods, leading to a minimal net impact on dedicated compute resource allocation.

## 2.3 Online Integration

During online serving, when a user request arrives the system performs the following steps -

1. **Parallel Retrieval**: Queries the standard online retrieval sources (Two Tower DNN, Content-KNN, Item-CF, rule-based sources) and simultaneously fetches the pre-computed top 200 ranked list for the user from the RADAR key-value store.
2. **Candidate Processing**: Candidates from standard online sources undergo the usual pre-ranking process. Candidates from the RADAR key-value store bypass the pre-ranking stage, as they have already been scored by a superior model.
3. **Merging & Final Ranking**: Candidates from all sources (post-pre-ranking for standard sources, direct for RADAR) are merged and deduplicated. This combined set is then fed into the final online ranking model for the ultimate re-ordering and selection before being presented to the user.

This integration ensures that high-quality, pre-vetted candidates from RADAR directly compete with candidates from traditional sources in the final ranking stage. RADAR, due to its offline nature and broader candidate evaluation, tends to surface more evergreen items and align with the user's long-term, stable interests. Conversely, traditional online retrieval sources are configured to adapt quickly to short-term user intent shifts and newly ingested items. Thus, RADAR and online sources act as complementary retrieval mechanisms, increasing the overall likelihood of surfacing a balanced mix of highly relevant and engaging content catering to both enduring preferences and immediate interests.

## 3 Offline Experiments

To evaluate the effectiveness of RADAR we conduct offline experiments on a large-scale dataset obtained from users' engagement on our video platform. We measure retrieval performance by instrumenting the standard recall@200 metric [3] using users' engagement on videos as ground truth. We compare RADAR against the following retrieval baselines:

1. **DNN:** a standard two-tower neural model[1] which maps users and items into a latent space combining collaborative filtering and content-based representations for each tower separately.

2. **Item-KNN** [4]. This is the standard item-based collaborative filtering method
3. **Content-KNN**: a nearest-neighbor baseline that retrieves items based solely on pretrained content embeddings using cosine similarity

We focus on three main research questions:

1. **RQ1: Retrieval Recall Performance** – How does RADAR perform on recall@200 compared to the baselines mentioned above?
2. **RQ2: Impact of Pool Size, Model Complexity** – Are the recall gains of RADAR primarily due to the much larger set of retrieved candidates or due to the stronger model used relative to pre-ranker model?
3. **RQ3: Performance Breakdown by User Segments** – Does RADAR perform equally well for all user cohorts?

## 3.1 RQ1: Retrieval Recall Performance

RADAR substantially outperforms the traditional retrieval baselines in Recall@200. As shown in Table 1, RADAR achieves about 16.5% recall@200, compared to 8.1% for the DNN Two Tower source and 7.2% for the Item-KNN source. We also observed that many of the items surfaced by RADAR have niche appeal that simple similarity-based methods failed to catch.

Table 1: Offline Recall@200 — RADAR vs. Baselines

| Retrieval Source | Recall@200 |
|---|---|
| DNN | 8.1% |
| Item-KNN | 7.2% |
| Content-KNN | 5.1% |
| RADAR | 16.5% |

## 3.2 RQ2: Retrieval Scaling vs. Model Scaling

To disentangle why RADAR excels, we ran ablation experiments varying the retrieved candidate pool size and model complexity. Table 2 summarizes the Recall@200 results for different configurations: (A) Base: RADAR configuration, (B) Scaled retrieved candidate pool size with simpler pre-ranker style model [9], (C) Online query candidate pool size with scaled up ranking model, (D) No scaling in candidate retrieved and ranking with simpler pre-ranker style model. Table 2 shows that both model scaling and retrieval scaling help increase RADAR performance, additionally we observe synergy between the two configs which further increase the performance of RADAR when used together.

Table 2: Recall@200 by Retrieval Scaling, Model Scaling

| Model Scaling | Retrieval Scaling | Recall@200 |
|---|---|---|
| Yes | Yes | 16.5% |
| No | Yes | 12.5% |
| Yes | No | 12.1% |
| No | No | 10.2% |

## 3.3 RQ3: User Performance Breakdown by Cohort

To further evaluate RADAR's performance, we segmented users into three activeness cohorts—Highly Active (daily active sessions), Moderately Active (engaging 2–3 times per week), and Dormant (at most 1 session every few weeks), based on their preceding 30-day interaction logs—and compared RADAR's Recall@200 against the strongest online baseline model (DNN) for each segment.

Table 3 summarizes recall@200 results for different user cohorts. Moderately active users benefit most likely because their usage frequency synchronizes with RADAR refreshes. Highly active users gain less: they exhaust cached lists quickly and their short-term interests shift faster than RADAR recommendations can be regenerated. No improvements are observed for Dormant users because RADAR cannot infer their interests and the cached list becomes stale, whereas an online DNN call still surfaces timely popular content for them

Table 3: Recall@200 by User Cohort

| User Cohort | Recall@200 (RADAR) | Recall@200 (DNN) |
|---|---|---|
| Highly Active | 16.2% | 8.2% |
| Moderately Active | 17.3% | 7.9% |
| Dormant | 6.5% | 6.9% |

## 4 Online Deployment

To validate RADAR in production, we ran a two-week A/B test on our video platform. The control kept the standard multi-stage funnel; the treatment injected the pre-computed top 200 RADAR candidates directly into the final ranker, bypassing online pre-ranking.

A key operational hurdle was candidate overlap, early experiments showed that many RADAR items were already being retrieved online, limiting incremental value. We therefore re-tuned the online retrieval generators to emphasize immediate, in-session users' intent and freshly uploaded items, letting RADAR specialize in users' stable long-term interests and evergreen content. After several tuning cycles we achieved ~60% unique candidates from RADAR significantly improving the incremental value.

With this configuration, the treatment delivered a +0.8% lift in our topline user engagement metric (statistically significant and correlated with long-term retention) and a +6% gain in unique item consumption, while keeping latency and system stability unchanged. This successful online validation confirms RADAR's effectiveness in leveraging asynchronous, offline computation to enrich the candidate pool for online ranking, thereby improving overall recommendation quality.